\documentclass[final,3p,times,twocolumn]{elsarticle}
 
\usepackage{amssymb}
\usepackage{amsmath}
\usepackage{graphics}
\usepackage{lineno}

\begin{document}

\begin{frontmatter}

\title{Production of the Doubly Charmed Baryons at the SELEX experiment -- The double intrinsic charm approach}
\author{Sergey Koshkarev}
\address{Institute of Physics, University of Tartu, Tartu 51010, Estonia}
\author{Vladimir Anikeev}
\address{Institute for High Energy Physics, Protvino 142281, Russia}

\begin{abstract}
The high production rate and $\langle x_{F} \rangle > 0.33$ of the
doubly charmed baryons measured by the SELEX experiment is not amenable
to perturbative QCD analysis. In this paper we calculate the production
of the doubly heavy baryons with the double intrinsic charm Fock states
whose existence is rigorously predicted by QCD. The production rate and
the longitudinal momentum distribution are both reproduced. We also
show that the production rates of the doubly charmed baryons and double
$J/\psi $ production observed by NA3 collaboration are comparable.
Recent experimental results are reviewed. The production cross section
of the doubly charmed baryons at a fixed-target experiment at the LHC is
presented.
\end{abstract}

\end{frontmatter}

\section{Introduction}%
\label{intro}

The SELEX measurements of the production of the doubly charmed baryons
at large $x_{F}$ are among the most intriguing and surprising results
in modern baryonic physics~\cite{SELEX2002,Mattson,SELEX2005}. The
SELEX experiment is a fixed-target experiment utilizing the Fermilab
negative charged beam at 600~GeV/c and positive beam at 800~GeV/c to
produce charm particles in a set of thin foil of Cu or in a diamond and
operated in the $x_{F} > 0.1$ kinematic region. The negative beam
composition was about 80\% $\Sigma^{-}$ and 20\% $\pi^{-}$. The positive
beam was 90\% protons.

In early 2000s the SELEX published first observation of 15.9 signal
over $6.1 \pm 0.5$ background events, at a mass of 3.52 GeV, of the
doubly charmed baryons in the charged decay mode $\Xi_{cc}^{+}\to
\Lambda_{c}^{+}K^{-} \pi^{+}$ from 1630 $\Lambda_{c}^{+}\to p K^{-}
\pi^{+}$ events sample~\cite{SELEX2002} which was previously used for
precision measurement of $\Lambda_{c}^{+}$
lifetime~\cite{SELEX2000,Kushnirenko}. Using same search strategy the
SELEX reported 20 signal events, at a mass of 3.76 GeV, of
$\Xi_{cc}^{++} \to \Lambda_{c}^{+}K^{-} \pi^{+} \pi^{+}$ decay mode
over 1656 $\Lambda_{c}^{+}\to p K^{-} \pi^{+}$ events
sample~\cite{Mattson}. In 2005 the SELEX collaboration published an
observation of 5.62 signal over $1.38 \pm 0.13$ background events, at a
mass of 3.52 GeV, of $\Xi_{cc}^{+}\to p D^{+} K^{-}$ decay mode from
1450 $D^{+} \to K^{-} \pi^{+} \pi^{+}$ decays to complement the
previous results~\cite{SELEX2005}. The SELEX measurements imply that
the lifetime of $\Xi_{cc}^{+}$ is less than 33 fs at 90\% confidence
level~\cite{SELEX2002}.

The production cross section has not been provided by the \mbox{SELEX}
collaboration. Still the production properties of $\Xi_{cc}^{+}$ and
$\Xi_{cc}^{++}$ can be compared to that of $\Lambda_{c}^{+}$ baryon:
\begin{eqnarray*}
R_{\Lambda_{c}^{+}} (\Xi_{cc}^{+}) = \frac{\sigma (\Xi_{cc}^{+})
\cdot Br(\Xi_{cc}^{+}\to \Lambda_{c}^{+}K^{-} \pi^{+})}{\sigma (
\Lambda_{c}^{+})}
\\
=\frac{N_{\Xi_{cc}^{+}}}{N_{\Lambda_{c}^{+}}} \cdot \frac{1}{\epsilon
_{+}} \approx 0.09
\end{eqnarray*}
and
\begin{eqnarray*}
R_{\Lambda_{c}^{+}} (\Xi_{cc}^{++}) = \frac{\sigma (\Xi_{cc}^{++})
\cdot Br(\Xi_{cc}^{++} \to \Lambda_{c}^{+}K^{-} \pi^{+} \pi^{+})}{
\sigma (\Lambda_{c}^{+})}
\\
=\frac{N_{\Xi_{cc}^{++}}}{N_{\Lambda_{c}^{+}}} \cdot \frac{1}{
\epsilon_{++}} \approx 0.045,
\end{eqnarray*}
where $N$ is number of events in the respective sample and the
reconstruction efficiencies of $\Xi_{cc}^{+}$ and $\Xi_{cc}^{++}$ are
$\epsilon_{+} \simeq 0.11$~\cite{SELEX2002} and $1/\epsilon_{++} \simeq
3.7$~\cite{Mattson} respectively. Such a high production rate with
$\langle x_{F} \rangle > 0.33$ and the relatively small mean transverse
momentum $\approx$1~GeV/c is not amenable to perturbative QCD
analysis~\cite{SELEX2002,Mattson}.

The production of states with two charm quarks with a high fraction of
a light hadron's momentum is unexpected if one adopts the conventional
assumption that heavy quarks can only arise from gluon splitting as in
DGLAP evolution.

However, QCD predicts another source of heavy quarks in the wavefunction
of a light hadron -- from diagrams where the heavy quarks are multiply
attached by gluons to the valence quarks~\cite{Brodsky1984,Franz}.
In this case, the frame-independent light-front wavefunction of the
light hadron has maximum probability when the Fock state is minimally
off-shell. This occurs when all of the constituents are at rest in the
hadron rest frame and thus have the same rapidity when the hadron is
boosted. Equal rapidity occurs when the light-front momentum fractions
of the Fock state constituents are proportional to their transverse
mass; i.e. when the heavy constituents have the largest momentum
fractions. This feature underlies the Brodsky, Hoyer, Peterson, and
Sakai (BHPS) model for the distribution of intrinsic heavy
quarks~\cite{Brodsky1980,Brodsky1981}.

Thus hadrons containing heavy quarks, such as the $\Lambda_{c}$, the
$J/\psi $, and even the doubly charmed baryons such as the $ccu$ or
$ccd$, can be produced in a hadronic collision with a high momentum
fraction of the beam momentum from the coalescence of the produced heavy
and valence quarks. The SELEX doubly charmed baryon results thus signify
a significant probability for the existence of Fock states such as
$| h_{l} c \bar{c} c \bar{c} \rangle $, where $h_{l}$ is light quark
content of the initial hadron.

\section{The doubly charmed baryons production cross section}\label{sec2}

In the BHPS model the wavefunction of a hadron in QCD can be represented
as a superposition of Fock state fluctuations, e.g. $| h \rangle
\sim | h_{l} \rangle + | h_{l} g \rangle + | h_{l} c \bar{c} \rangle
\ldots{}$, where $ h_{l}$, as above, is light quark content. When the
projectile interacts with the target, the coherence of the Fock
components is broken and the fluctuation can hadronize. The intrinsic
charm Fock components are generated by virtual interactions such as
$gg \to c \bar{c}$ where gluon couple to two or more projectile valence
quarks. The probability to produce such $ c \bar{c}$ fluctuations scales
as $\alpha_{s}^{2} (m_{c}^{2})/m_{c}^{2}$ relative to leading-twist
production.

Following \cite{Brodsky1980,Brodsky1981,Vogt1995} a general
formula for the probability distribution of an $n$-particle intrinsic
charm Fock state as a function of $x_{i}$ and transverse momentum
$\vec{k}_{T,i}$ can be written as:
%
\begin{equation}
 \frac{dP_{ic(c)}}{\prod_{i=1}^n dx_i d^2 k_{T,i}} \propto \frac{\delta \big(\sum_{i=1}^n \vec{k}_{T,i} \big)\delta \big( 1 - \sum_{i=1}^n x_i \big)}{\big( m_h^2 -  \sum_{i=1}^n m_{T,i}^2 / x_i \big)^2},
\end{equation}
where $m_{T,i}$ denotes $\sqrt{m_{i}^{2} + k_{T,i}^{2} }$ and
$m_{h}$ is mass of the initial hadron. Let us denote the probability of
$| h_{l} c \bar{c} \rangle $ and $| h_{l} c \bar{c} c \bar{c} \rangle
$ Fock states as $P_{ic}$ and $P_{icc}$. In this paper we will also
simplify the formula with replacement $m_{T,i}$ with the effective mass
$\hat{m}_{i} = \sqrt{m_{i}^{2} + \langle k^{2}_{T,i} \rangle }$ and
neglect the mass of the light quarks. This model assumes that the vertex
function in the intrinsic charm wavefunction is relatively slowly
varying; the particle distributions are then controlled by the
light-cone energy denominator and phase space. The Fock states can be
materialized by a soft collision in the target which brings the state
on shell. The distribution of produced open and hidden charm states will
reflect the underlying shape of the Fock state wavefunction.

\subsection{The double intrinsic charm approach}\label{sec2.1}

We assume that all of the doubly charmed baryons are produced from
$|h_{l} c \bar{c}c \bar{c} \rangle $ Fock states. In the quark-hadron
duality approximation the probability to produce a $\Xi_{cc}$ is
proportional to the fraction of $c c$ production below threshold mass
$m_{th} = m_{D} + \Delta m$~\cite{KiselevUFN}, where $m_{D}$ is
$D$-meson mass and $\Delta m \simeq 0.5\hbox{--}1~\mbox{GeV}$. The fraction of
$c c$ pairs can be written as:
%
\begin{equation}
f_{cc/h} \simeq \int_{4m_{c}^{2}}^{m_{th}^{2}} dM_{cc}^{2} \frac{dP
_{icc}}{dM_{cc}^{2}}
\,\,
\bigg/ \int_{4m_{c}^{2}}^{s} dM_{cc}^{2} \frac{dP_{icc}}{dM_{cc}^{2}}.
\end{equation}
To obtain the fraction ratio of $c c$ pairs into $\Xi_{cc}$ baryons we
have to isolate color-antitriplet states, the fraction ratio of the
doubly charmed baryons is
%
\begin{equation}
\label{eq:aaa}
f_{\Xi_{cc}/h} \approx s_{c} \cdot f_{cc/h} \, ,
\end{equation}
where the $s_{c}$ is the color-antitriplet factor. The $cc$ pair has
$3 \times 3 = 9$ color components, 3 color-antitriplet, and 6
color-sixtet. Assuming that $cc$ are unpolarized in the color space in
the double intrinsic charm Fock state, there is $1/3$ probability for
the color-antitriplet possibility. Finally, we get $s_{c} \simeq 2
\times 1/3$. Let us remind the reader that some of c-quarks could
produce open charm states so we need to interpret $f_{\Xi_{cc}/h}$ as
the upper limit.

If we take $m_{c} = 1.5~\mbox{GeV}$ the value of $f_{\Xi_{cc}/p} \approx 0.6$.
This model also predicts $f_{\Xi_{cc}/\Sigma^{-}} \simeq f_{\Xi_{cc}/p}$
that is comparable with the SELEX data~\cite{SELEX2002}.

There is a simple connection between the intrinsic charm cross section
and the inelastic one~\cite{Vogt1995,Vogt,Ingelman}
%
\begin{eqnarray}
\nonumber
\sigma_{icc} = \frac{P_{icc}}{P_{ic}} \cdot \sigma_{ic}
\\
\label{eq:bbb}
\sigma_{ic} = P_{ic} \cdot \sigma^{in} \frac{\mu^{2}}{4 \hat{m}_{c}
^{2}} \approx 3 \cdot 10^{-5} \sigma^{in} \, ,
\end{eqnarray}
where $\mu^{2} \approx 0.2~\mbox{GeV}^{2}$ denotes the soft interaction scale
parameter; $P_{ic} \simeq 0.3\hbox{--}2\mbox{\%}$ (see~\cite{Duan} and
references therein). In the Ref.~\cite{Vogt1995} it is found that
for proton $P_{icc} \approx 20 \% \cdot P_{ic}$. In our calculation we
use the following approximation~\cite{Vogt,Ingelman}:
%
\begin{equation}
\label{eq:ccc}
\sigma_{ic} = 0.1 \cdot \sigma_{pQCD} (c \bar{c}).
\end{equation}
The normalization is fixed to be the same as Eq.~\ref{eq:bbb} at
$\sqrt{s} = 20\hbox{--}40~\mbox{GeV}$.

Combining Eqs.~{\eqref{eq:aaa}, \eqref{eq:bbb}
and \eqref{eq:ccc}{\ref{eq:aaa}, \ref{eq:bbb} and
\ref{eq:ccc}} we may expect the upper limit of the production cross
section of the doubly charmed baryons to be:
\begin{eqnarray*}
\sigma_{icc}(\Xi_{cc}) \simeq f_{\Xi_{cc}/p} \,
\frac{P_{icc}}{P_{ic}} \cdot 0.1 \cdot \sigma_{pQCD} \approx 7 \cdot
10^{4} \, \mbox{pb},
\end{eqnarray*}
where $\sigma_{pQCD} (c \bar{c}) \approx \sigma (gg \to c \bar{c})
\simeq 5.8 \times 10^{6}~\mbox{pb}$ is the charm production cross section at
600~GeV/c beam momentum, where most of statistics was collected,
calculated with CalcHEP Monte-Carlo tool~\cite{CalcHep}.

\subsection{The intrinsic charm approach}\label{sec2.2}

The intrinsic charm production cross section of the doubly charmed
baryons can be written as follows:
\begin{eqnarray*}
\sigma_{ic}(\Xi_{cc}^{+}) = \int dx_{1} dx_{2} f_{g} (x_{1}, \mu ) f
_{c} (x_{2}, \mu ) \hat{\sigma }(x_{1}, x_{2}),
\end{eqnarray*}
where $f_{g,c} (x, \mu )$ is the gluon~\cite{CTEQ6L} or intrinsic
charm~\cite{Pumplin} distribution functions, $x$ is the ratio of
the parton momentum to the momentum of the hadron and $\mu $ is the
energy scale of the interaction. Explicit form of $\hat{\sigma }(gc
\to \Xi_{cc}^{+})$ can be found in~\cite{ChangPRD}. In the SELEX
case these calculations have been done in Ref.~\cite{Chang2006}:
\begin{eqnarray*}
\sigma_{ic}(\Xi_{cc}^{+}) \simeq 102 \, \mbox{pb}.
\end{eqnarray*}
The value is relatively small and can be neglected.

\subsection{The total production cross section}\label{sec2.3}

The charm quark fragmentation into the doubly charm baryon and the
perturbative approaches give too small a contribution and can be also
neglected so the total production cross section of the doubly charmed
baryons at the SELEX experiment will be:
\begin{eqnarray*}
\sigma (\Xi_{cc}) \approx \sigma_{icc}(\Xi_{cc}) \approx 7 \cdot 10^{4}
\, \mbox{pb}.
\end{eqnarray*}

It is interesting to estimate of the doubly charmed baryon production
at a fixed-target experiment at the LHC~\cite{Brodsky2013} with
$\sqrt{s} \simeq 115~\mbox{GeV}$. Following the method described above the
production cross section of the doubly charmed baryons will be:
\begin{equation*}
\sigma_{icc}(\Xi_{cc}) \simeq f_{\Xi_{cc}/p} \cdot \frac{P_{icc}}{P
_{ic}} \cdot 3 \cdot 10^{-5} \sigma^{in} \approx 1.5 \times 10^{6} \,
\mbox{pb} ,
\end{equation*}
where the value of $f_{\Xi_{cc}/p} \approx 0.56$ and
$\sigma^{in} (\sqrt{s} \simeq 115~\mbox{GeV}) \approx 28.4~\mbox{mb}$ \cite{Block}.
It is two order of magnitude bigger than predicted in
ref.~\cite{Chen2014} with the single intrinsic charm approach.

\section{The shape of $P_{icc} (\Xi_{cc})$ as a function of
$x_{F}$}
\label{kin}

As we already mentioned (see Sec.~\ref{intro}) the large mean
$x_{F}$ and small mean transverse momentum is not amenable to
perturbative QCD analysis. The $x_{F}$ distribution of $\Xi_{cc}$
baryons can be written as:
%
\begin{equation}
\frac{dP_{icc}(\Xi_{cc})}{dx_{F}} = \int \prod_{i=1}^{n} dx_{i} \frac{dP
_{icc}}{dx_{1} ... dx_{n}} \times \delta (x_{\Xi } - x_{c} -x_{c}).
\end{equation}
The mean value of $x_{F}$ is
%
\begin{equation}
\label{eq:mean}
\langle x_{F} \rangle = \int dx_{F} \, x_{F} \frac{dP_{icc}(\Xi_{cc})}{dx
_{F}}.
\end{equation}
Integrating Eq.~\ref{eq:mean} over the
SELEX kinematic region, $x_{F} > 0.1$, we find $\langle x_{F} \rangle
\approx 0.33\hbox{--}0.34$ that is in agreement with the SELEX data,
$x_{F} \sim 0.33$ for $\Xi_{cc} ^{+}\to \Lambda_{c}^{+}K^{-} \pi^{+}$
decay~\cite{SELEX2002}. In his PhD thesis, Mattson provide the $x_{F}$
distribution of the $\Xi_{cc}^{++}$ candidates into
$\Lambda_{c}^{+}K^{-} \pi^{+} \pi^{+}$ decay mode~\cite{Mattson}.
Integrating Eq.~\ref{eq:mean} over the
region where data presents, $x_{F} > 0.175$ (see
Fig.~\ref{fig:shape}), we find $\langle x_{F} \rangle \approx
0.36$ that also agrees with the data. The relatively small transverse
momentum also is a sign of the intrinsic charm mechanism.

\begin{figure}[h]
\label{fig:shape}
\includegraphics[scale=0.45] {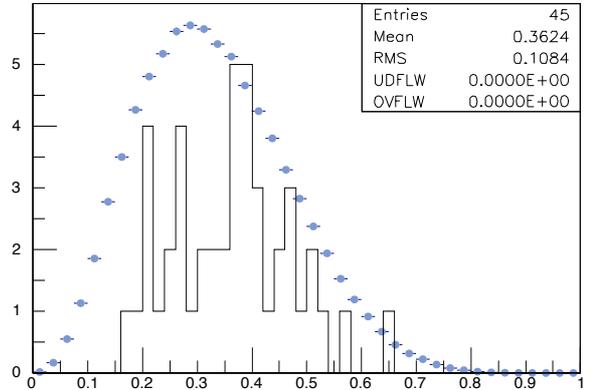}
\caption{The solid line histogram shows number of the SELEX $\Xi_{cc}^{++}$
candidates as a function of $x_{F}$~\cite{Mattson}. The dotted histogram
represents calculation of this distribution in the double intrinsic charm
model.}
\end{figure}

\section{Solving mystery of the SELEX result}%
\label{solving}

As we noted above, the SELEX collaboration did not provide the doubly
charmed baryons production cross section but we are still able to
compare it to the production properties of the $\Lambda_{c}^{+}$
baryons. Let us remind the reader the measured ratios: $R_{\Lambda
_{c}^{+}}^{\mathit{exp}} (\Xi_{cc}^{+}) = \sigma (\Xi_{cc}^{+}) \cdot Br(\Xi
_{cc}^{+}\to \Lambda_{c}^{+}K^{-} \pi^{+}) / \sigma (\Lambda_{c}^{+})
\approx 0.09$ and $ R_{\Lambda_{c}^{+}}^{\mathit{exp}} (\Xi_{cc}^{++}) =
\sigma (\Xi_{cc}^{++}) \cdot Br(\Xi_{cc}^{++} \to \Lambda_{c}^{+}K
^{-} \pi^{+} \pi^{+}) / \sigma (\Lambda_{c}^{+}) \approx 0.045$. In the
leading order perturbative QCD the production cross section of the
$\Lambda_{c}^{+}$ baryons can be approximated as:
\begin{eqnarray*}
\sigma (\Lambda_{c}^{+}) \approx \sigma (gg \to c \bar{c}) \cdot f(c
\to \Lambda_{c}^{+}).
\end{eqnarray*}
The SELEX search strategy of the doubly charmed baryons requires
minimum value of $x_{\Lambda_{c}^{+}} >
0.15$~\cite{SELEX2002,Mattson,SELEX2000,Kushnirenko}. Assuming that
$x_{\Lambda_{c}^{+}} \sim x_{c}$ and using CalcHEP Monte-Carlo tool
find $ \sigma (gg \to c\bar{c})|_{x_{c} > 0.15} \sim 3 \cdot
10^{5}~\mbox{pb}$, fragmentation ratio $f(c \to \Lambda_{c}^{+}) =
0.071 \pm 0.003~(\mbox{exp}.)\pm 0.018~(\mbox{br}.)$~\cite{BaBarLambda}
so the production cross section of the $\Lambda_{c}^{+}$ baryons at the
SELEX experiment will be:
\begin{eqnarray*}
\sigma_{pQCD}(\Lambda_{c}^{+})|_{x_{\Lambda_{c}^{+}} > 0.15} \approx
\sigma (gg \to c\bar{c})|_{x_{c} > 0.15} \cdot f(c \to \Lambda_{c}
^{+})
\\
\approx 2.1 \cdot 10^{4} \, \mbox{pb}.
\end{eqnarray*}
Using the branching ratios predicted by J. Bjorken $Br(\Xi_{cc}^{+}
\to  \Lambda_{c}^{+}K^{-} \pi^{+}) =  0.03$ and $Br(\Xi_{cc}^{++} \to
\Lambda_{c}^{+}K^{-} \pi^{+} \pi^{+}) = 0.05$~\cite{BaBar}, one can
obtain the ratio of the production cross sections:
%
\begin{equation}
\label{eq:ddd}
R^{th}_{\Lambda_{c}^{+}} = \frac{\sigma (\Xi_{cc}) \cdot Br(\Xi_{cc}
^{+(+)} \to \Lambda_{c}^{+}K^{-} \pi^{+} (\pi^{+}))}{\sigma (\Lambda
_{c}^{+})|_{x_{\Lambda_{c}^{+}} > 0.15}} \sim 0.15.
\end{equation}
However this result is not really accurate. Playing with parameters we
can change both doubly charmed baryons and charm production cross
sections in wide enough range. The most important thing about the ratio
(\ref{eq:ddd}) is that it has the same order of magnitude as the
measured ones against a few order of magnitude gap another predictions
provide~\cite{KiselevUFN,ChangPRD,Chang2006}.

The relatively high production rate of $c \bar{c} c \bar{c}$ states to
charm is not a unique feature of the SELEX experiment. The double
$J/\psi $ production properties measured by the NA3
experiment~\cite{NA31982,NA31985} have many similar features: the
high $\sigma (\psi \psi )/\sigma (\psi ) = (3 \pm 1) \times 10^{-4}$
rate, large $x_{\psi \psi }$ and small average transverse momentum,
$p_{T,\psi \psi } = 0.9 \pm 0.1~\mbox{GeV}/\mbox{c}$. It is interesting to compare the
SELEX result with the NA3 data on the double $J/\psi $ production. The
NA3 experiment is a beam dump experiment at CERN utilizing antiprotons,
protons, pions and kaons at 150, 200 and 280~GeV/c to produce charm
particles with incident on hydrogen and platinum targets in the
$x_{F} > 0$ kinematic region. The most informative data the NA3
collaboration present is the double $J/\psi $ production with
$\pi^{-}$ beam at 280~GeV/c. It is not possible to compare the SELEX and
the NA3 data directly but we are able to compare the following ratios,
where $R$ denotes $\sigma ({c \bar{c} c \bar{c}}) / \sigma (c \bar{c})$:
\begin{eqnarray*}
R_{SELEX} \sim R_{\Lambda_{c}^{+}}^{exp} \times \frac{f(c \to \Lambda
_{c}^{+})}{f_{\Xi /p}} \approx (0.8 \pm 0.2) \times 10^{-2}
\end{eqnarray*}
and
\begin{eqnarray*}
R_{NA3} \sim \frac{\sigma (\psi \psi )}{\sigma (\psi )} \times \frac{f
_{J/\psi }}{f_{\psi /\pi }^{2}} \approx 2 \times 10^{-2},
\end{eqnarray*}
where $f_{\psi /\pi } \approx 0.03$~\cite{Vogt1995} and
$f_{J/\psi } \approx 0.06$~\cite{CERN2004}. Therefore, as we can
see, the NA3 data complements the hight production rate at the SELEX
experiment.

\section{Review of Belle and LHCb recent results}%
\label{subsec:review}

The Belle experiment~\cite{Belle} presented the upper limit on the
$\sigma (e^{+}e^{-} \to \Xi_{cc}^{+}X)$ is 82--500~fb for the decay mode
with the $\Lambda_{c}^{+}$ at $\sqrt{s} = 10.58~\mbox{GeV}$ using
$980~\mbox{fb}^{-1}$. The most realistic
calculations~\cite{KiselevUFN,Kiselev94} of the upper limit cross
section predict $\sigma (\Xi_{cc}^{+}) \simeq 35 \pm 10~\mbox{fb}$ what turns
out to be at least twice as less as the given limit.

Another recent result from the LHCb experiment~\cite{LHCb} provides the
upper limits at 95\% C.L. on the ratio $\sigma (\Xi_{cc} ^{+}) \cdot
Br(\Xi_{cc}^{+}\to \Lambda_{c}^{+}K^{-} \pi^{+})/\sigma (
\Lambda_{c}^{+})$ to be $1.5 \times 10^{-2}$ and $3.9 \times 10^{-4}$
for lifetimes 100 fs and 400 fs respectively, for an integrated
luminosity of $0.65~\mbox{fb}^{-1}$. It is compared with result from
Ref.~\cite{KiselevUFN,Chang2006,ChangPRD,Gunter} $\sim
10^{-4}\hbox{--}10^{-3}$. However, the LHCb did not reach the lifetime
measured by the SELEX experiment yet. Moreover, the LHCb analysis
requires that $\Lambda_{c}^{+}$ candidates have to be significantly
displaced from the primary vertex so this requirement potentially cuts
down most of the signal region. The contribution from the double
intrinsic charm is suppressed due to LHCb experiment kinematics.
Assuming that the hadron identification efficiency for pions and kaons
is degraded above 100~GeV/c \cite{RICH} (such that when raised to the
fourth power it is negligible), and making the naive assumption that
momentum is split evenly between all final-state tracks, the analysis
loses sensitivity around $p(\Xi_{cc})=500~\mbox{GeV}/\mbox{c}$, i.e. $x_{F}=0.14$.

\section{Summary}\label{sec6}

The experimental results (see Sec.~\ref{intro}) and theoretical
predictions (see Sec.~\ref{solving}) on the production properties of
the $\Xi_{cc}$ in the \mbox{SELEX} experiment have the same order of magnitude
accuracy. The predicted mean Feynman-x values (see Sec.~\ref{kin})
agree with the experimental data. The NA3 collaboration result on the
double $J/\psi $ production strongly complements the SELEX data. We
would like to specially point out the fact that unexpectedly high
production rate of $\Xi_{cc}$ baryons is due to the kinematics features
of the SELEX experiment, and could not be described by the production
mechanism only. We also find that the doubly intrinsic charm approach
will be the leading production mechanism of the doubly charmed baryons
at high Feynman-$x$ at a future fixed-target experiment at the LHC.

\subsection*{Acknowledgments}
We would like to thank Prof. Stanley Brodsky for pointing out the
intrinsic charm mechanism, explanation of its nature and very useful
discussions. We would also like to thank Dr. Matthew Charles and Dr.
Yury Shcheglov for very useful comments on the acceptance of the LHCb
detector, Prof. Stefan Groote for useful comments on the manuscript
structure and Vladislav Sukhmel for proofreading the early manuscript
draft.

This work was supported by the under Grant No. IUT2-27.


\begin{thebibliography}{00}

\bibitem{SELEX2002}
M.~Mattson {\sl et al.} (SELEX~Collaboration), Phys. Rev. Lett. 89,  112001 (2002),~ArXiv:hep-ex/0208014.

\bibitem{Mattson}
M.~Mattson, Ph.D. thesis, Carnegie Mellon University, 2002.

\bibitem{SELEX2005}
A.~Ocherashvili {\sl et al.} (SELEX~Collaboration), Phys. Lett. B{\bf628}, 12-24 (2005),~ArXiv:hep-ex/0406033.

\bibitem{SELEX2000}
A.~Kushnirenko {\sl et al.}, Phys. Rev. Lett. {\bf86}, 5243 (2001),~ArXiv:hep-ex/0010014.

\bibitem{Kushnirenko}
	A.~Kushnirenko, Ph.D. thesis, Carnegie Mellon University, 2000.
	
\bibitem{Brodsky1984}
	S.~Brodsky {\sl et al.}, CNUM: C84-06-23 (1984).

\bibitem{Franz}
	M.~Franz {\sl et al.}, Phys. Rev. D{\bf62}, 074024 (2000).

\bibitem{Brodsky1980}
	S.~Brodsky {\sl et al.}, Phys. Lett. B{\bf93}, 451 (1980).

\bibitem{Brodsky1981}
	S.~Brodsky {\sl et al.}, Phys. Rev. D{\bf23}, 2745 (1981).

\bibitem{Vogt1995}
	R, Vogt and S. Brodsky, Phys. Lett. B349, 569-575 (1995), ArXiv:hep-ph/9503206.

\bibitem{KiselevUFN}
	V.~Kiselev and A.~Likhoded, Phys. Ups. {\bf45}, 455 (2002),~ArXiv:hep-ph/0103169.

\bibitem{Vogt}
	R. Vogt, S. Brodsky and P. Hoyer, Nucl. Phys. B360, 67 (1991).

\bibitem{Ingelman}
	G,~Ingelman, M.~Thunman, Z.Phys. C73, 505-515 (1997), ArXiv:hep-ph/9604289.

\bibitem{Duan}
	Shaorong Duan, C. S. An, B. Saghai, Phys. Rev. D93, 114006 (2016), ArXiv:1606.02000.

\bibitem{CalcHep}
	A.~Belyaev, N.~Christensen, A.~Pukhov,~ArXiv:1207.6082.

\bibitem{CTEQ6L}
	J.~Pumplin {\sl et al.}, JHEP {\bf07}, 012 (2002),~ArXiv:hep-ph/0201195.

\bibitem{Pumplin}
	J.~Pumplin, Phys. Rev. D73, 114015 (2006), ArXiv:hep-ph/0508184.

\bibitem{ChangPRD}
	C.-H.~Chang {\sl et al.}, Phys. Rev. D73, 094022 (2006),~ArXiv:hep-ph/0601032.

\bibitem{Chang2006}
	C.-H.~Chang {\sl et al.}, J.Phys.G34,845 (2007),~ArXiv:hep-ph/0610205.

\bibitem{Brodsky2013}
	S.J. Brodsky, F. Fleuret, C. Hadjidakis, J.P. Lansberg, Phys. Rep. 522, 239-255 (2013).

\bibitem{Block}
  M.M.~Block and F.~Halzen,  Phys.\ Rev.\ D {\bf 86}, 014006 (2012), ~ArXiv:1205.5514.
	
\bibitem{Chen2014}
	Gu Chen {\sl et al.} ,Phys. Rev. D89, 074020 (2014), ArXiv:1401.6269.	

\bibitem{BaBarLambda}
	B.~Aubert {\sl et al.} (BaBar~Collaboration), Phys. Rev. D{\bf75}, 012003 (2007),~ArXiv:hep-ex/0609004.

\bibitem{BaBar}
	B.~Aubert {\sl et al.} (BaBar Collaboration),Phys. Rev. D{\bf74}, 011103 (2006),~ArXiv:hep-ex/0605075.

\bibitem{NA31982}
	J.~Badier {\sl et al.}, Phys. Lett. B{\bf114}, 457 (1982).

\bibitem{NA31985}
	J.~Badier {\sl et al.}, Phys. Lett. B{\bf158}, 85 (1985).
	
\bibitem{CERN2004}
	M.~Mangano, H.~Satz and U.~Wiedemann, eds., CERN Yellow report, CERN-2004-009 (2004) .

\bibitem{Belle}
	Y.~Kato {\sl et al.} (Belle Collaboration), Phys. Rev. D{\bf89}, 052003 (2014),~ArXiv:1312.1026.

\bibitem{Kiselev94}
	V.~Kiselev, A.~Likhoded, M.~Shevlyagin, Phys. Lett. B{\bf332}, 411-414 (1994),~ArXiv:hep-ph/9408407.

\bibitem{LHCb}
	R.~Aaij {\sl et al.} (LHCb Collaboration), JHEP {\bf12}, 090 (2013),~ArXiv:1310.2538.

\bibitem{Gunter}
	D. G\"{u}nter and V. Saleev, Phys. Atom Nucl. 65, 299-304 (2002), ArXiv:hep-ph/0104173.

\bibitem{RICH}
	M.~Adinolfi {\sl et al.} (LHCb Collaboration), Eur. Phys. J. C{\bf73}, 2431 (2013), ArXiv:1211.6759. 

\end{thebibliography}
\end{document}